\documentclass[useAMS,usenatbib]{mn2e}
\usepackage{graphicx}
\usepackage{amssymb}
\usepackage{lscape}
\usepackage{ulem}
\usepackage{txfonts}

\def\kms{km~s$^{-1}$}

\def\si2{Si\,{\sc ii}}
\def\mg2{Mg\,{\sc ii}}
\def\fe2{Fe\,{\sc ii}}
\def\al2{Al\,{\sc ii}}
\def\zn2{Zn\,{\sc ii}}
\def\c2s{C\,{\sc ii}$^{\star}$}
\def\hkpc{$h_{70}^{-1}$ kpc}

\def\dv{$\Delta$V}

\title[AGN in galaxy pairs]
{Galaxy pairs in the Sloan Digital Sky Survey - IV: Interactions trigger AGN.}

\author[Ellison et al.] {Sara L. Ellison$^1$, David R. Patton$^2$,
J. Trevor Mendel$^1$, Jillian M. Scudder$^1$. \\
$^1$ Department of Physics and Astronomy, University of Victoria, Victoria, British Columbia, V8P 1A1, Canada.\\
$^2$ Department of Physics \& Astronomy, Trent University, 
1600 West Bank Drive, Peterborough, Ontario, K9J 7B8, Canada.
}

\begin{document}

\maketitle

\begin{abstract}
Galaxy-galaxy interactions are predicted to cause gas inflows leading
to enhanced nuclear star formation.  This prediction is borne out
observationally, and also supported by the gas-phase metallicity dilution
in the inner regions of galaxies in close pairs.  In this paper we
test the further prediction that the gas inflows lead to enhanced
accretion onto the central supermassive black hole, triggering
activity in the nucleus.  Based on a sample of 11,060 Sloan Digital
Sky Survey galaxies with a close companion ($r_p < 80$ \hkpc,
$\Delta$V $<$ 200 \kms), we classify active galactic nuclei (AGN)
based either on emission line ratios or on spectral classification
as a quasar.  The AGN fraction in the close pairs sample is compared
to a control sample of 110,600 mass- and redshift-matched control
galaxies with no nearby companion.  We find a clear increase in the
AGN fraction in close pairs of galaxies with projected separations $<$
40 \hkpc\ by up to a factor of 2.5 relative to the control sample (although
the enhancement depends on the chosen S/N cut of the sample). The
increase in AGN fraction is strongest in equal mass galaxy pairings,
and weakest in the lower mass component of an unequal mass pairing.
The increased AGN fraction at small separations is accompanied by an
enhancement in the number of `composite' galaxies whose spectra are
the result of photoionization by both AGN and stars.  Our results
indicate that AGN activity occurs (at least in some cases) well before
final coalescence and concurrently with ongoing star formation.
Finally, we find a marked increase at small projected separations of
the fraction of pairs in which \textit{both} galaxies harbour AGN.  We
demonstrate that the fraction of double AGN exceeds the expected
random fraction, indicating that some pairs undergo correlated nuclear
activity.  We discuss some of the factors that have led to conflicting
results in previous studies of AGN in close pairs.  Taken together
with complimentary studies, we favour an interpretation where interactions
trigger AGN, but are not the only cause of nuclear activity.
\end{abstract}


\section{Introduction}

There are several lines of evidence that indicate that the build-up of
stellar mass in galactic bulges is closely linked to the evolution of
the central black hole (e.g. Haehnelt, Natarajan \& Rees 1998;
Richstone et al. 1998).  Gravitational torques have been shown to be
effective at funneling gas to the centres of galaxies to feed both
star formation and accretion onto the central black hole (e.g. Haan et
al. 2009) and there is a remarkably tight relationship between the
stellar properties of the bulge and the mass of the black hole
(Magorrian et al 1998; Ferrarese \& Merritt 2000; Gebhardt et
al. 2000). The similarities in the shape of the redshift evolution of
both star formation rates and the frequency of active galactic nuclei
(AGN) indicate a common physical driver between the two processes.
There is considerable observational evidence linking AGN activity with
star formation, such as bluer colours in AGN hosts (Silverman et
al. 2009) and correlations between star formation activity and AGN
power (Cid-Fernandes et al. 2001; Choi, Woo \& Park 2009).  In turn,
both observations and simulations suggest that feedback from the AGN
itself may quench the star formation (Kauffmann et al. 2003a; Di
Matteo et al. 2005; Schawinski et al. 2007, 2009; Bundy et al 2008;
van de Voort et al. 2011; Kaviraj et al. 2011).

The obvious culprit for building both the central stellar mass and
feeding the black hole (and hence increasing its activity) is a ready
supply of gas for fuel and galaxy-galaxy mergers represent a natural
mechanism for providing this gas in wholesale quantities (Kauffmann \&
Haehnelt 2000; Cattaneo et al. 2005; Di Matteo et al. 2005;Springel,
Di Matteo \& Hernquist 2005; Wild et al. 2007; Hopkins et al. 2008).
Many authors have found evidence for mergers associated with AGN hosts
(e.g. Surace et al. 1998; Canalizo \& Stockton 2001; Sanchez \&
Gonzalez-Serrano 2003; Jahnke et al. 2004; Smirnova et al. 2006;
Combes et al. 2009; Villar-Martin et al. 2010, 2011), leading to an
often cited connection between the merger process and AGN triggering.
The discovery of double AGN also lend circumstantial support to a
merger origin (e.g. McGurk et al. 2011 and references therein).
However, although some studies claim a connection between nuclear
activity and the presence of close companions (Dahari 1984; Keel et
al. 1985; Rafanelli et al. 1995; Koss et al. 2010), other studies have
claimed that there is statistically no difference in the percentage of
galaxies with companions between active and inactive galaxies (Schmitt
2001; Coldwell \& Lambas 2006; Grogin et al. 2005).  Similarly, there
are claims that AGN hosts are no more tidally distorted than inactive
galaxies (Dunlop et al. 2003; Gabor et al. 2009; Cisternas et
al. 2011; Kocevski et al. 2011), but counter claims also exist
(e.g. Koss et al. 2010; Ramos Almeida 2011a,b).  Various biases may
contribute to these discrepant results, including sample size,
wavelength-dependent dust extinction, image depth, adequate definition
of a control sample, automated versus visual classification and
distinction between Type I and Type 2 AGN (e.g. Darg et al. 2009;
Dultzin et al. 2010; Koss et al. 2010; Liu et al. 2011a,b; Ramos
Almeida et al. 2011b).  Nonetheless, the null results have led some
authors to conclude that there is actually little evidence for a
connection between merging and AGN triggering (e.g. Grogin et
al. 2005; Gabor et al. 2009; Cisternas et al. 2011).  These
conclusions potentially undermine one of the cornerstones of our
modern paradigm of galaxy evolution: the triggering of AGN through
merger events.

A natural way to test the connection between mergers and AGN is to use
close pairs of galaxies which can be classified as either star-forming
or AGN-dominated on the basis of their emission line ratios.  Close
pairs have been unequivocally demonstrated to have enhanced star
formation rates (Kennicutt et al., 1987; Barton, Geller \& Kenyon
2000; Lambas et al 2003; Alonso et al. 2004; Nikolic, Cullen \&
Alexander 2004; Woods, Geller \& Barton 2006; Woods \& Geller 2007;
Ellison et al. 2008, 2010; Patton et al. 2011; Liu et al. 2011b) and
evidence for gas inflows, as indicated by diluted interstellar medium
metallicities (Kewley et al. 2006a; Ellison et al. 2008; Michel-Dansac
et al. 2008; Kewley et al. 2010; Rupke et al. 2010; Scudder et al. in
preparation).  Similar enhancements in star formation rates (SFRs) 
and low gas phase
metallicities have been reported for other classes of galaxies that
are likely to be post-mergers, such as luminous infrared and
`lopsided' galaxies (Rupke, Veilleux \& Baker 2008; Reichard et
al. 2009). It is therefore perhaps surprising that the results for AGN
fractions in close pairs are considerably more controversial.  For
example, Alonso et al. (2007), Woods \& Geller (2007) and Rogers et
al. (2009) all find higher AGN fractions in samples of close pairs,
relative to a control sample of field galaxies.  Conversely, Li et
al. (2006, 2008) and Ellison et al. (2008) and Darg et al (2009) do
not find a statistically significant difference in AGN fractions.

One of the issues that hampers the observational association between
mergers and AGN ignition is the expected time delay between clear
signs of merging (e.g. well identified galactic components and strong
tidal features), the presence of triggered star formation and,
finally, the activity in the nucleus.  This complication was
highlighted by Storchi-Bergmann et al. (2001) in their study of
Seyfert 2 galaxies.  The majority of the Seyferts with close
companions show signs of recent nuclear star formation, leading
Storchi-Bergmann et al. to suggest that the pure Seyfert spectrum
emerges after the main starburst has faded.  A similar evolutionary
scenario was suggested by Haan et al. (2008), whereby low luminosity
AGN precede Seyferts after a merger, based on high fractions of the
former, but few of the latter in a sample of galaxies with disturbed
HI disks.  Delays are also implied by the intermediate colours and
spectral characteristics of AGN hosts, indicating at least $\sim$ 100
Myr between the decline of star formation and black hole activity
(Schawinski et al. 2009).  Indeed, by the time galaxies have migrated
from their peak star-forming days into the `green valley' and/or show
signs of strong AGN, their morphologies do not exhibit any residual
signs of interactions (Cisternas et al. 2011; Mendez et al. 2011).
Delays of a few hundred megayears between the epochs of peak star
formation and peak black hole accretion are also inferred from
simulations (Wild, Heckman \& Charlot 2010;  Hopkins 2011).  Although gas
inflow leads to triggered star formation after the first pericentric
passage, the highest enhancements are seen when the galaxies finally
merge.  If coalescence also represents the peak in AGN activity, this
may further contribute to the difficulty in connecting AGN with galaxy
pairs.

In this paper, we use a large sample (Section \ref{sample_sec}) of
close galaxy pairs to return to the question of AGN fractions in
galaxy-galaxy interactions.  One novelty in the current study is the
application of a variety of AGN classifications (Section
\ref{diag_sec}) which permit different contributions from AGN and star
formation.  The varying strictness of these diagnostics allows us to
obtain some measure of how dominant the AGN contribution is.  This
will be useful if our sample includes a significant fraction of
galaxies that are just transitioning between their star-forming and
AGN phases, the so-called composite population.  Most previous works
have used a single sample of close pairs (with separations typically
$<$ 30 \hkpc) and calculated a single AGN fraction to be compared
with the field (e.g. Alonso et al. 2007; Ellison et al. 2008; Darg et
al. 2009).  However, one of the benefits of our large sample size is
that we can investigate the AGN fraction \textit{as a function of
projected separation} and as a function of pair mass ratio (Section
\ref{frac_sec}).  This has only been attempted once before by Woods \&
Geller (2007) with a sample around 1/3 the size of ours.  Finally, we
will investigate whether there is any evidence of simultaneous
triggering of AGN (Section \ref{double_sec}).  In particular we will
quantify whether the fraction of double AGN pairs is consistent with
the expectation from random occurrence, or whether there is a
component of correlated AGN pairs.

\section{Sample selection}\label{sample_sec}

The sample of pairs used in this work differs significantly from that
used in previous papers in this series (Ellison et al. 2008, 2010;
Patton et al. 2011) in several respects.  Previously, the pairs sample
was constructed from only the SDSS galaxy (specclass=2) sample.
However, in this work, we are interested in the AGN contribution, so
we have extended our search for pairs to also include objects
classified as QSOs (specclass=3) which have been classified as
galaxies from the SDSS imaging.  For clarity, we will refer to objects
with specclass values of either 2 or 3 as `galaxies'.  The parent
sample of galaxies from which we will select the pairs is restricted
to the SDSS legacy area (i.e. excluding the Segue footprint) with
extinction corrected $r$-band Petrosian magnitudes in the range $14.0
< m_r < 17.77$. Our previous works have used stellar masses calculated
from SDSS photometry made available by the Max Planck Institut fur
Astrophysik/Johns Hopkins University (MPA/JHU)
collaboration\footnote{http://www.mpa-garching.mpg.de/SDSS/}.
However, in Simard et al. (2011), we have shown that the SDSS
photometry becomes unreliable for galaxies with close companions (see
also Figure 10 in Patton et al. 2011).  Simard et al. (2011)
recomputed the magnitudes for 1.2 million SDSS galaxies using improved
sky subtraction and object definition techniques, combined with
morphological decompositions.  These improvements dramatically
improved the quality of photometric measurements in close pairs
(e.g. Figure 11 of Simard et al. 2011).  We therefore re-compute
stellar masses using the updated photometry of Simard et al. (2011).

The masses are derived using a simple relation between g-r colour and
mass-to-light ratio (M/L) derived from MPA/JHU data catalogues, which
we parametrise as a double power-law.  M/L measurements in the MPA/JHU
catalogues were derived from fits to the SDSS $ugriz$ {\tt model}
photometry to a suite of synthetic SEDs spanning a broad range of
star-formation histories (Kauffmann et al. 2003b; Salim et al. 2007),
and are corrected for the influence of emission lines on the SDSS
photometry.  Model SEDs were constructed using {\sc galaxev} (Bruzual
\& Charlot 2003) with a Chabrier (2003) initial mass function, and
galaxies were assumed to be well described by the combination of an
underlying, exponentially declining star-formation history and
stochastically-sampled bursts.  Simard et al. (2011) show that their
magnitudes are on average 0.05 to 0.1 mag brighter that the SDSS {\tt
model} magnitudes and, more importantly for this work, are less prone
to deblending errors at small galaxy--galaxy separations (see also
Patton et al. 2011).  We therefore adopt our updated masses throughout
this work, although note that our results are qualitatively similar if
we use the MPA/JHU stellar masses.  

We impose a minimum redshift of 0.01 to avoid non-cosmological values
and a maximum value of 0.20 above which the sample becomes
increasingly incomplete.  The final criterion for inclusion in the
parent sample is that the SDSS redshift confidence parameter must have
a value of $z_{conf} > 0.7$, i.e. the confidence level of the redshift
is greater than 70 per cent (our results are unaffected by increasing
this cut to higher confidence levels).

From the parent sample we select pairs and higher order
multiples\footnote{94 per cent of our sample consists of pairs,
approximately 6 per cent are triples and less than 1 percent are
higher order multiples.} with a companion within 80 \hkpc, \dv\ $<$
200 \kms\ and with stellar mass ratios 0.1 $< M_1/M_2< 10$.  Fibre
collisions lead to a high incompleteness at separations $<$ 55
arcseconds which biases the mass and redshift distribution of close
pairs (Ellison et al. 2008; Patton \& Atfield 2008).  We therefore
follow Ellison et al.  (2008) and exclude a random 67.5\% of pairs
with $\theta >$ 55 arcseconds to yield a pairs sample with unbiased
selection as a function of separation.  Before the cull is
implemented, there is a clear trend towards larger masses at small
separations.  Since there is a strong dependence of AGN fraction on
mass, this artificially introduces an increased AGN fraction at small
separations.  After the cull, the distrubution of masses and redshifts
with projected separation is flat.  There are 11,060 galaxies in the
full pairs sample of which 46 are classified as QSOs.

A control sample is constructed by iteratively matching galaxies in
both mass and redshift from a pool of galaxies with no close companion
within 80 \hkpc\ and 10,000 \kms.  The control sample is vital since
the AGN fraction depends sensitively on galactic properties such as
mass, luminosity, colour, morphology and concentration (Kauffmann et
al. 2003b; Best et al. 2005; Choi, Woo \& Park 2009; Ellison et
al. 2008).  Matching in mass largely mitigates these differences,
as well as dealing with aperture bias (the varying covering fraction
as a function of redshift).
The matching procedure finds the best simultaneous match in mass
and redshift to each galaxy (i.e. both members of the pair are matched
separately) and calculates the Kolmogorov-Smirnov (KS) probability
that the masses and redshifts of the control and pairs are drawn from
the same distribution.  If the KS probability is above 30\%, the
matching is repeated up to the $N^{th}$ best match.  We are able
to match 10 controls to each galaxy in a pair before the KS test fails,
yielding a control sample of 110,600 galaxies.  We keep track of which
controls are matched to which pair galaxy so that we can include only the
relevant controls for different subsamples of pairs.

\section{AGN diagnostics}\label{diag_sec}

\begin{figure}
\centerline{\rotatebox{270}{\resizebox{6cm}{!}
{\includegraphics{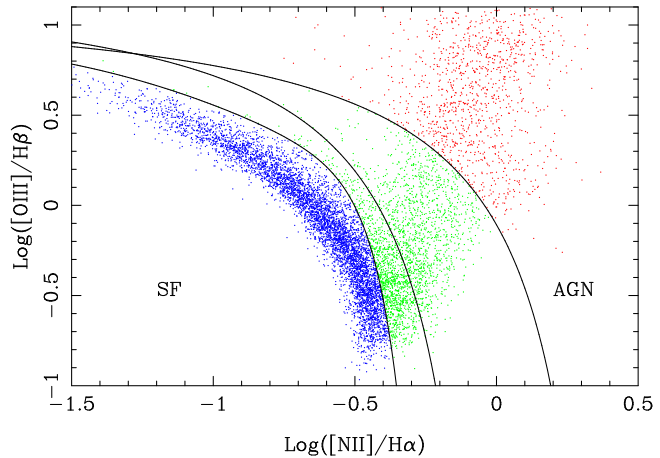}}}}
\caption{\label{bpt} The BPT diagram for control galaxies with strong
emission lines.  Black curves indicate the demarcation lines given by
K01 (upper line), K03 (middle line) and S06 (lower line), see
equations \ref{K01_BPT_eqn} to \ref{S06_BPT_eqn}.  Galaxies are
colour-coded for easy distinction: star-forming (blue), composite
(green) and AGN (red). QSOs not present in the MPA/JHU line flux
catalogues are excluded from this Figure.}
\end{figure}

\subsection{Background}

\begin{figure}
\centerline{\rotatebox{0}{\resizebox{8cm}{!}
{\includegraphics{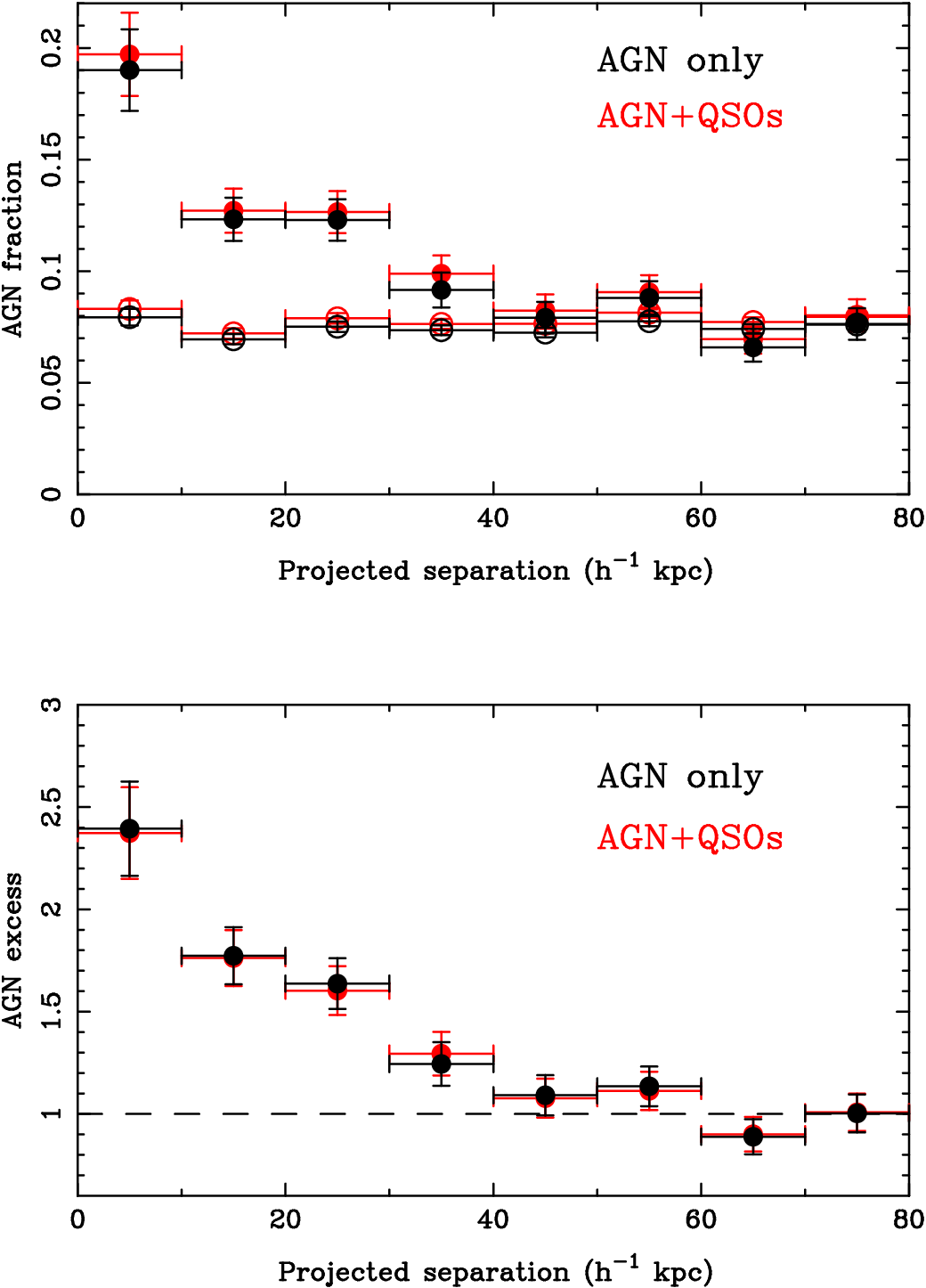}}}}
\caption{\label{frac_rp} Top panel: The fraction of AGN in close
galaxy pairs (filled points) as a function of projected separation.
Open points show the AGN fraction for the control sample, where the
projected separation corresponds to the pair galaxy to which it is
matched.  Black symbols consider only galaxies classified as AGN from
the S06 BPT diagnostic.  Red symbols include the specclass=3 (QSO)
objects. Lower panel:  The AGN excess is determined from the ratio of the
AGN fraction in the pairs relative to the matched control, as a function
of projected separation.  AGN are up to 2.5 times more common in close galaxy
pairs than the control (independently of whether QSOs are included or not).}
\end{figure}

Since their early use in distinguishing spectra dominated by AGN
(Baldwin, Phillips \& Terlevich 1981; Veilleux \& Osterbrock 1987),
line ratio diagrams have evolved into a standard tool for categorizing
the dominant source of ionizing radiation in a galaxy (Dopita et
al. 2000; Kewley et al. 2001, 2006b; Kauffmann et al. 2003a; Stasinska
et al. 2006; Groves et al. 2006; Cid-Fernandes et al. 2010).  The
principle behind these diagnostic diagrams is to distinguish the
excitation mechanism (photoionization by hot stars, 
AGN or shock heating).  The various mechanisms result in a different
ionization structure in the interstellar gas, and changes in the
relative sizes of the ionized, partly ionized and neutral media.  The
inner zone is highly ionized, hosting species such as [OIII].
The outer zone is the primary location of the lower ionization species
such as [OII] and some of the [NII].  Forming a partially ionized zone
requires a relatively hard spectrum as the decreasing cross section of
hydrogen (and other species) means that only the high energy photons
penetrate.  This means that the partially ionized zone is prevalent in
the presence of an AGN and shocks, but virtually absent with stellar
spectra. [NII], [OI] and [SII] are all formed in the partially ionized
zone, and can be boosted further by collisional
excitation.  The differences in structure and ionization balance are
encapsulated in a single ionization parameter, $q$: the ratio of
hydrogen ionizing photon flux per unit area to the local number
density of hydrogen atoms.  In terms of the choice of line ratios used
to distinguish the ionizing source,
preference is usually given to those that are relatively close in
wavelength in order to avoid a significant dependence on internal
extinction properties.

The most commonly used combination of line ratios is [NII]/H$\alpha$
versus [OIII]/H$\beta$, often simply referred to as the BPT diagram
after Baldwin, Phillips \& Terlevich (1981).  Although galaxies
dominated by star-formation and AGN broadly distinguish themselves as
two `wings' on the BPT diagram, the precise demarcation is not clear.
Using photoionization models, Kewley et al. (2001, K01) proposed a
classification scheme that extends into the AGN wing of the BPT diagram.
The Kewley et al. (2001) classification includes the most extreme
models of starburst galaxies in terms of metallicity and star
formation rate.  However, galaxies classified as star-forming by
the Kewley (2001) system may contain up to $\sim$ 20\% contribution
from AGN (Stasinska et al. 2006).  Stasinska et al. (2006, S06) also model
a range of ionization parameters and metallicities, but with updated
model atmospheres, resulting in a more stringent
demarcation between AGN and star-forming galaxies.  Between these
extremes, Kauffmann et al. (2003a, K03) have proposed an empirical
distinction between AGN and HII-region dominated galaxies, based on a
sample of $\sim$ 23,000 galaxies from the SDSS.

\subsection{Application to the SDSS sample}

\begin{figure}
\centerline{\rotatebox{270}{\resizebox{6cm}{!}
{\includegraphics{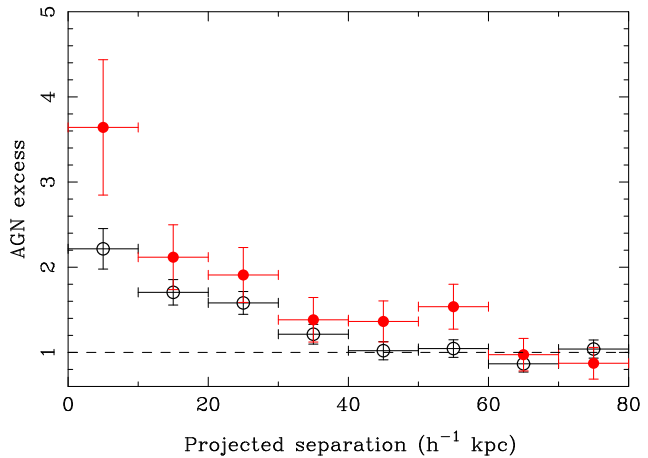}}}}
\caption{\label{frac_z} The AGN excess (fraction of AGN in the pairs
relative to the fraction of AGN in the control sample) is plotted
as a function of projected separation.  Black open symbols show
galaxies at $z<0.1$ and red filled points show galaxies at $z\ge0.1$.}
\end{figure}

\begin{figure*}
\centerline{\rotatebox{270}{\resizebox{12cm}{!}
{\includegraphics{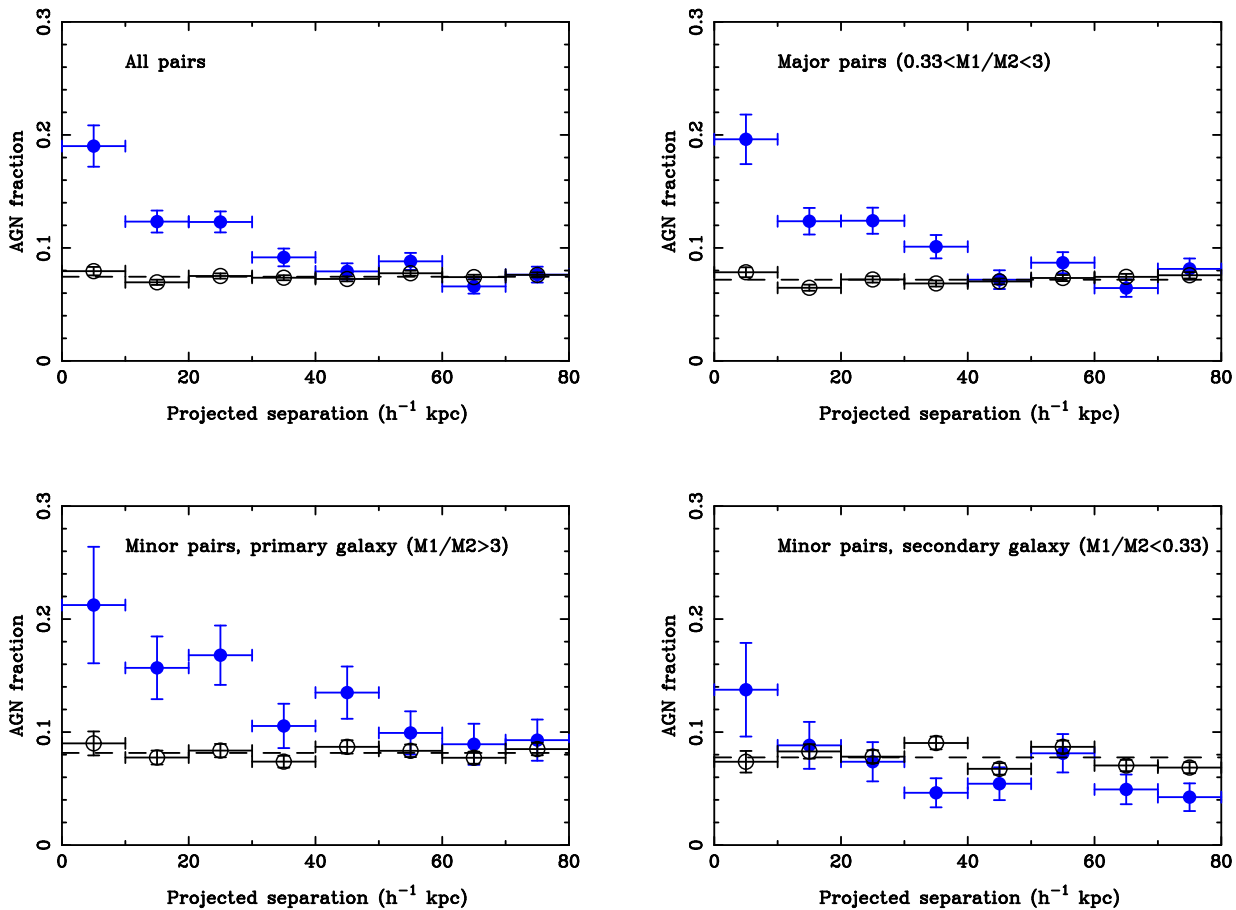}}}}
\caption{\label{3mratio} The fraction of pair galaxies (solid points)
classified as
AGN according to the S06 diagnostic as a function of projected
separation.  Open points show the AGN fractions for the control galaxies 
matched to the pairs in each bin of projected separation.  The panels 
show the full pairs sample (top left), only the
approximately equal mass (major) pairs (top right), the greater (bottom
left) and lower mass (bottom right) in unequal  mass (minor) pairs.  }
\end{figure*}

Galaxies in our sample are classified as AGN based on the ratios of
strong emission lines, or based on their spectral classification as
QSOs (specclass=3).  Emission line fluxes are taken from the MPA/JHU
catalogues.  The catalogue fluxes are already corrected for underlying
stellar absorption and Galactic extinction.  Galaxies not present in
the MPA/JHU catalogues are discarded (this includes some specclass=3
QSOs, most notably those with large Balmer line widths).  Fluxes are
further corrected for internal extinction using the ratio of
H$\alpha$/H$\beta$ and an SMC extinction curve (Pei 1992).  The
minimum criterion for AGN classification is the detection at $>5
\sigma$ of H$\beta$ $\lambda 4861$, [OIII] $\lambda$ 5007, H$\alpha$
$\lambda 6563$ and [NII] $\lambda 6584$.  There are also cuts on the
flux error and continuum error of the emission lines that clip
spurious measurements, see Scudder et al. (in preparation) for full
details.

In this paper, we will investigate all three of the common BPT
classification schemes: Kewley et al (2001), Kauffmann et al. (2003a)
and Stasinska et al. (2006).  The distinction between AGN and star-forming
galaxies for the 3 schemes are given in equations \ref{K01_BPT_eqn}
(K01), \ref{K03_BPT_eqn} (K03) and \ref{S06_BPT_eqn} (S06).

\begin{equation}\label{K01_BPT_eqn}
\log (\rm{[OIII]}\lambda 5007 / \rm{H}\beta) = \frac{0.61}{\log(\rm{[NII]}\lambda6584/\rm{H}\alpha)-0.47} + 1.19
\end{equation}

\begin{equation}\label{K03_BPT_eqn}
\log (\rm{[OIII]}\lambda 5007 / \rm{H}\beta) = \frac{0.61}{\log(\rm{[NII]}\lambda6584/\rm{H}\alpha)-0.05} + 1.30
\end{equation}

\begin{eqnarray}\label{S06_BPT_eqn}
\log (\rm{[OIII]}\lambda 5007 / \rm{H}\beta) = (-30.787 + 1.1358\log(\rm{[NII]}\lambda6584/\rm{H}\alpha) \nonumber \\
+ 0.27297(\log(\rm{[NII]}\lambda6584/\rm{H}\alpha)^2))tanh(5.7409\log(\rm{[NII]}\lambda6584/\rm{H}\alpha)) \nonumber \\
- 31.093
\end{eqnarray}

Figure \ref{bpt} shows how these boundaries map on to the standard
emission line diagram for our control sample.
We consider galaxies classified by the K01 scheme as AGN, to be `pure'
AGN and galaxies classified by S06 as star-forming to be `pure'
star-forming.  Galaxies classified as AGN by S06 and star-forming by
K01 are considered as `composite' (e.g. Ho, Filippenko \& Sargent
1993) objects as they may contain contributions from both
AGN and HII region spectra.  Galaxies may also
reside in the composite zone if they have particularly high
star-formation rates, high metallicities or harbour shock-excited gas.
AGN may themselves be divided into Seyfert galaxies and those that exhibit
weaker, low ionization emission lines, the so-called `low ionization
nuclear emission line region' objects (LINERs, Heckman 1980).
Although the excitation mechanism for LINERs is itself
ambiguous, it is likely to originate from a non-stellar process
(Ho et al. 1993).  In this paper, we do not consider these
finer AGN classifications or the effects of shock excitation.
Although the latter may be present in mergers, the SDSS spectra
can not easily diagnose its presence (Rich et al. 2010, 2011).

\section{The AGN fraction in galaxy pairs}\label{frac_sec}

We begin simply, by using the BPT diagram to classify galaxies as AGN
or non-AGN.  For this exercise, we first use the S06 diagnostic to
distinguish between the two galaxy classes, as it includes even a
modest amount of nuclear activity. Galaxies with a QSO classification
are also included in the AGN sample.  Figure \ref{frac_rp} shows the
AGN fraction of pairs as a function of projected separation ($r_p$).
Note that we quote the AGN fraction of the full pairs sample, which
contains both emission line galaxies that can be classified as AGN on
the BPT diagram, and quiescent galaxies with no strong emission lines.
For comparison, we show results both with and without the minority of
galaxies with a QSO classification.    There is a steady increase
towards small projected separation in the AGN fraction for pairs with
$r_p< 40$ \hkpc.  The control galaxies show no such dependence on
$r_p$, where the projected separation for the control galaxies
reflects the value of the pairs to which they are matched.  This
result qualitatively supports the findings of Alonso et al. (2007) who
found a 10\% increase in AGN fraction at $r_p < 25$ kpc (see also
Woods \& Geller 2007 and Rogers et al. 2009).  However, we note that
the actual fractional increase in AGN is not comparable between ours
and previous works, whose samples contain only emission line galaxies
(see Section \ref{comp_sec} for further discussion).  The lower panel
of Figure \ref{frac_rp} shows the ratio of the AGN fraction in pairs
relative to the control, i.e. the excess of AGN in the pairs.  At the
closest separations ($r_p < 10$ \hkpc) AGN are 2.5 times more common
in pairs than in the control sample.  The inclusion of QSOs makes a negligible
impact on these statistics.

By splitting the sample by redshift at a value of 0.1, we show in
Figure \ref{frac_z} the AGN excess in two redshift bins.  The $z>0.1$
sample has a slightly higher AGN excess at all values of $r_p$, but
the significance is $<1 \sigma$ except at the smallest separations
($r_p < 10$ \hkpc).  Any increase in AGN fraction in the higher
redshift bin is unlikely to be due to the effect of covering fraction for two
reasons.  First, the AGN excess is relative to a mass- and
redshift-matched control sample, so that we expect the covering
fraction to be approximately the same between a given pair galaxy and
its controls.  Second, the higher covering fraction of high redshift
galaxies should lead to a lower fractional contribution from AGN,
rather than an enhanced AGN fraction.  Although Figure \ref{frac_z} may
therefore indicate some dependence of AGN fraction on redshift, the
redshift interval probed by our sample is too narrow to allow a
rigourous investigation of the redshift dependence of the merger-AGN
connection.

In Figure \ref{3mratio} we plot the AGN fraction of paired galaxies as
a function of projected separation for different mass ratios.  Once
again, the AGN fractions of the control galaxies matched to the pairs
in each bin of projected separation are shown for comparison.  The
major (equal mass) pairs show an enhancement that is very similar to
the full sample.  However, in minor (unequal mass) pairs, the more
massive galaxies show a strong increase in the AGN fraction, whereas
the signal is marginal in the less massive companion.  This mass
dependence is the inverse of what has been previously observed for
triggered star formation in close pairs, where the primary galaxy
shows no enhancement, but the secondary galaxy does (Woods \& Geller
2007), as predicted by simulations (Cox et al. 2008).  None the less,
our AGN results agree with those of Woods \& Geller (2007) who report
a possible (1--2 $\sigma$) increased AGN fraction for the primary
galaxy in a minor merger.  On the other hand, in a study of dual AGN,
Liu et al. (2011b) find that both [OIII] luminosities and black hole
accretion rates are higher in the less massive than the more massive
component of a minor pair.

It is interesting to note that in the secondary galaxy of unequal mass
pairs in our sample at projected separations $r_p > 30$ \hkpc\ there
is a hint that the AGN fraction may actually be lower than in the mass
matched control.  The lack of enhanced AGN fraction in the secondary
galaxies is presumably not due to the lack of fuel being supplied to
the galactic centre.  Simulations predict significnt torques in the
secondary galaxy (Cox et al. 2008) and SFRs are indeed observed to
be elevated (Woods \& Geller 2007).  The relative trends of enhanced
SFRs and AGN fractions in pairs of different mass ratios may therefore
provide some clues to the timescales on which the two phases dominate.

\begin{figure}
\centerline{\rotatebox{0}{\resizebox{8cm}{!}
{\includegraphics{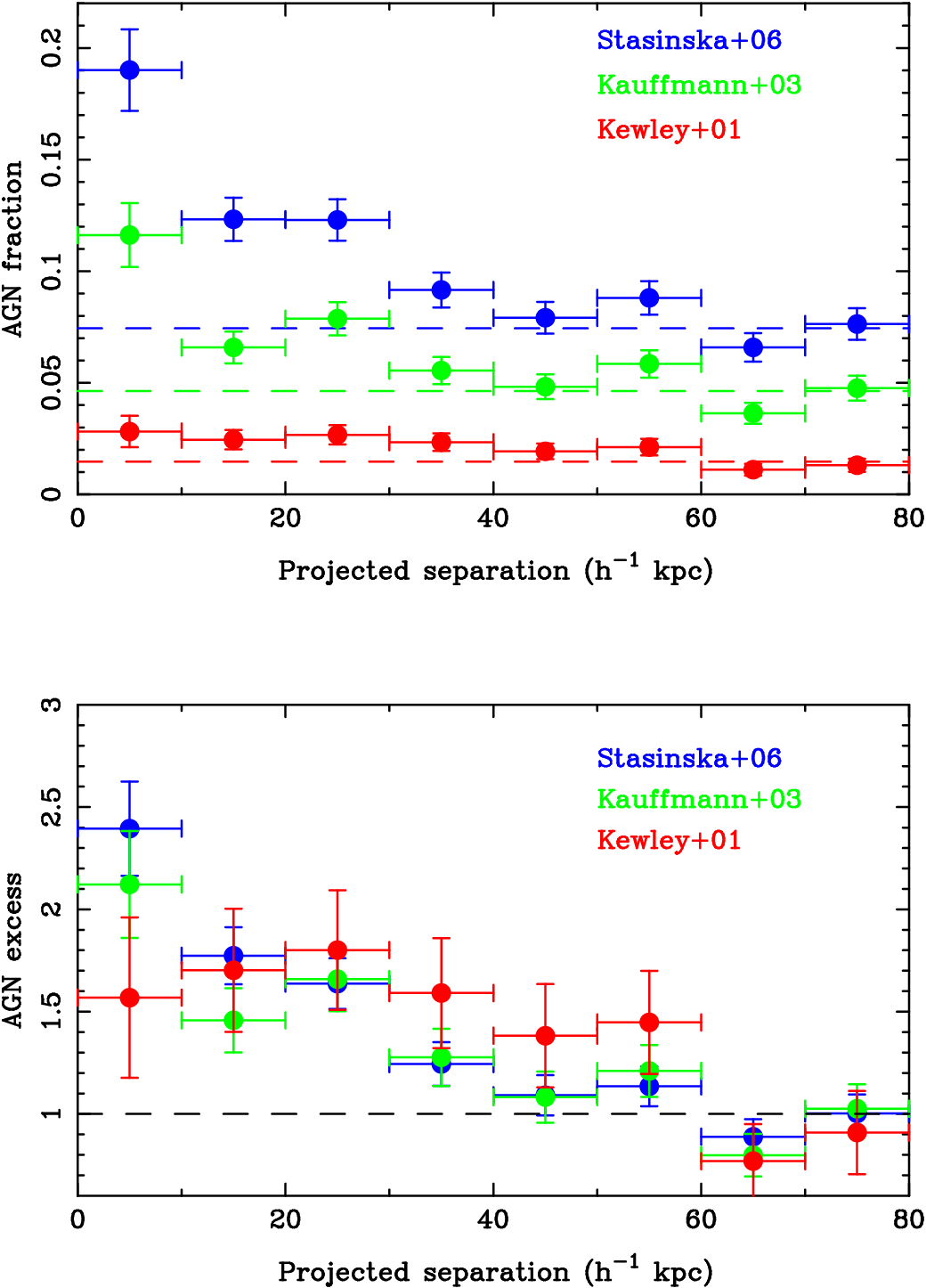}}}}
\caption{\label{3diag} The fraction of pair galaxies classified as AGN
according to the K01 (red), K03 (green) and S06 (blue) diagnostic as a
function of projected separation.  The dashed lines indicate the
median AGN fraction in the control sample.  The top panel shows the
absolute fraction and the lower panel shows the excess relative to
the control.  QSOs not present in the
MPA/JHU line flux catalogues are excluded from this Figure.}
\end{figure}

In Figure \ref{3diag} we investigate how the choice of AGN diagnostic
(K01, K03 or S06) influences the AGN fraction as a function of
$r_p$ (this figure necessarily excludes QSOs which do not have
BPT classifications).  As expected, the overall AGN fractions increase
from K01 to S06.  The lower panel of Figure \ref{3diag} shows the AGN
excess relative to the control sample. Despite some small bin-to-bin
differences, the overall trend for all three diagnostics is an increasing
AGN fraction towards smaller projected separations with a typical
over-abundance of AGN at $r_p < 10$ \hkpc\ of a factor of about two.

\begin{figure}
\centerline{\rotatebox{0}{\resizebox{8cm}{!}
{\includegraphics{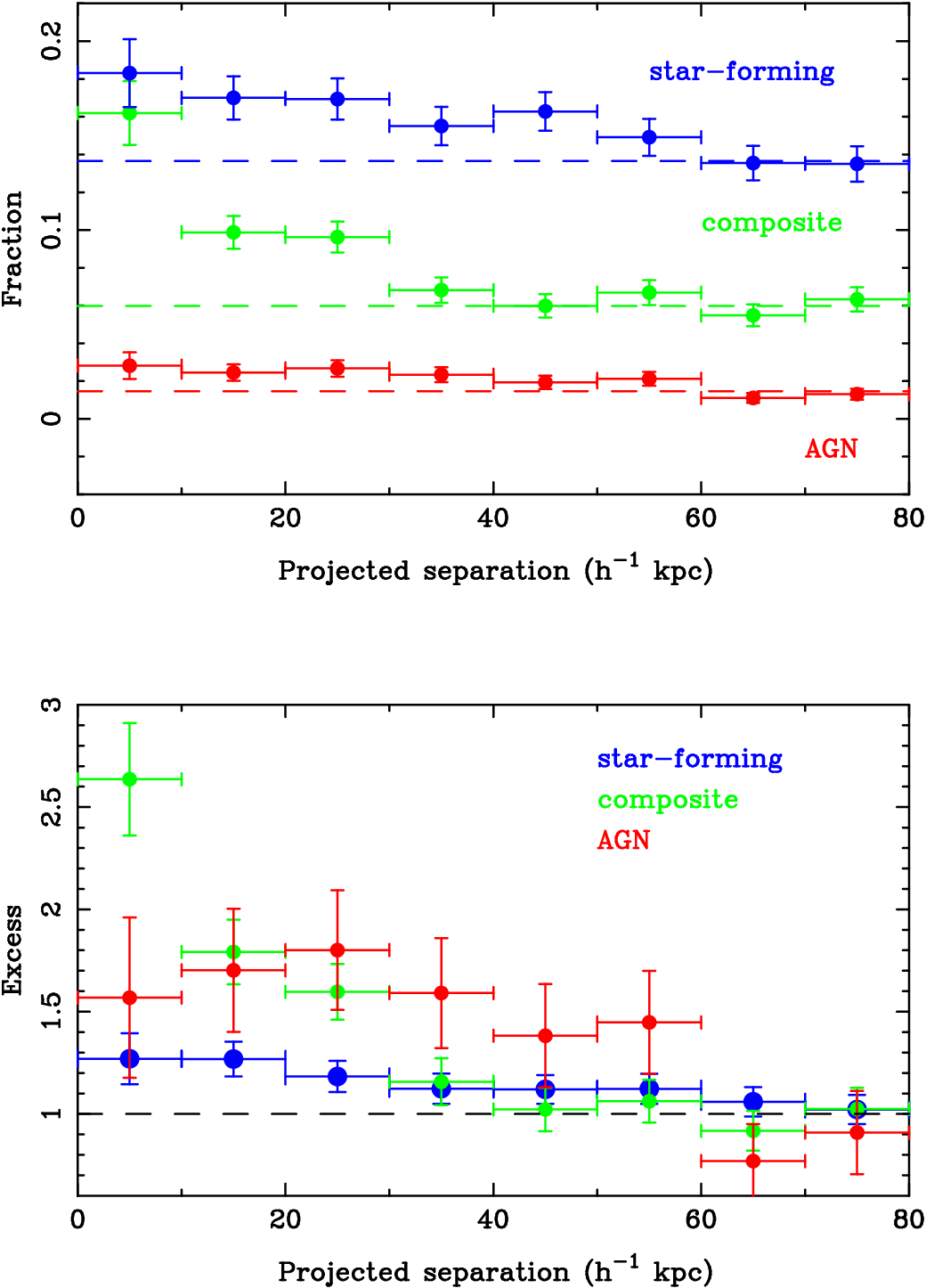}}}}
\caption{\label{tran_frac} The fraction of pair galaxies classified as
AGN (red), composite (green) and star-forming (blue) galaxies as a 
function of projected
separation.  The dashed lines indicate the median AGN fraction in the
control sample.  QSOs not present in the
MPA/JHU line flux catalogues are excluded from this Figure.}
\end{figure}

Due to the varying contributions of AGN permitted by the different
diagnostics, the three classifications can be combined to classify
three populations. `Pure' AGN and `pure' star-forming galaxies are
based on the K01 and S06 demarcations.  Emission line galaxies located
between the K01 and S06 lines are considered to be `composite' in
nature.  Figure \ref{tran_frac} shows the fraction and excesses
(relative to the control) of the pure AGN, pure star-forming and
composite galaxies.  All three categories show an excess at small
separations, indicating that the overall emission line fraction is
increasing in small separation pairs.  Presumably the emission line
population is being fed by relatively quiescent galaxies with sufficient
residual gas that an interaction can spur them into (star formation or
AGN) activity.  Interestingly, although one of the best known properties of
close galaxy pairs is their high star formation \textit{rates}, the actual
number of galaxies in the pure
star-forming category shows the most modest enhancement, increasing by
only 20\%.  The `pure' AGN fraction increases by 50\%, but the
composite galaxies, which have significant contributions from both AGN
and star formation show the largest excess at $r_p < 10$ \hkpc, by a
factor of 2.5.  Large fractions of composite (also sometimes referred
to as low luminosity AGN or transition objects) have been previously
noted in pairs studies by Pastoriza, Donzelli \& Bonatto (1999) and
Focardi, Zitelli \& Marinoni (2008). 

\section{Double AGN}\label{double_sec}

\begin{figure}
\centerline{\rotatebox{0}{\resizebox{8cm}{!}
{\includegraphics{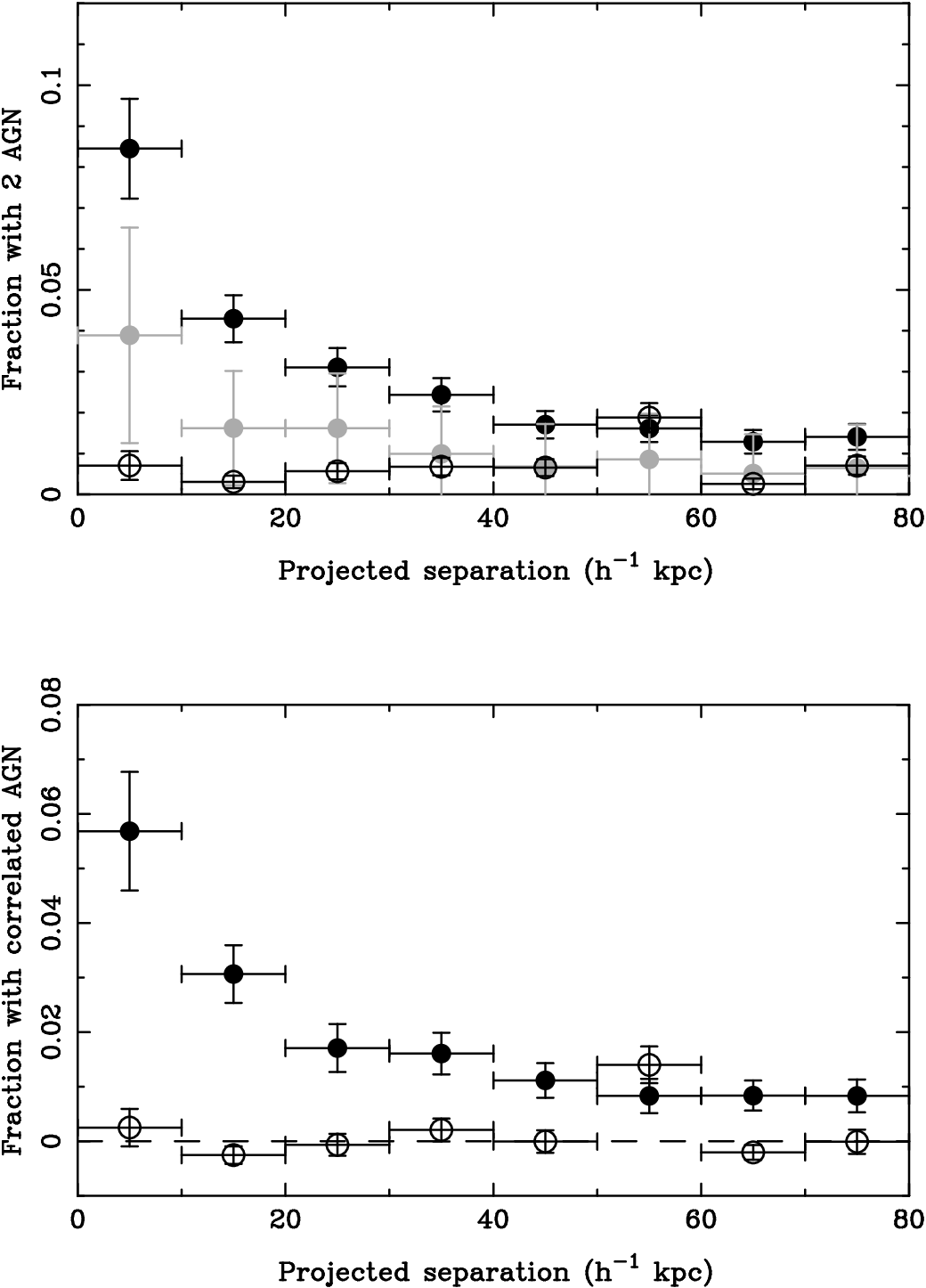}}}}
\caption{\label{x_jk} Top panel: The fraction of pairs with two AGN as
a function of projected separation (filled black points).  The grey
points indicate the expected (random) double AGN fraction based on the
values in Figure \ref{frac_rp}.  There is a clear excess of observed
double AGN relative to the random case.  A sample of control pairs is
constructed by taking the best matched control (out of ten) of each
paired galaxy and pairing it up with the best matched control to the
companion galaxy.  The fraction of double AGN in the control pairs is
shown with open symbols.  Lower panel: The fraction of correlated AGN
pairs ($xf$, see Table \ref{xeqns_tab}) as a function of projected
separation.  Pairs are shown with solid symbols, control galaxies are
shown with open symbols.  The enhanced fraction of correlated double
AGN (relative to the control) indicate that both galaxies in the pair
are undergoing synchronized activity.}
\end{figure}

In Figure \ref{x_jk} we investigate the fraction of close pairs in
which both of the galaxies are classified as an AGN.  In order
to calculate the double AGN in a control sample, we make fake pairs
by taking the best matched control galaxy of each galaxy in a pair. 
In Figure
\ref{x_jk} we elect to use the S06 diagnostic, but the same
qualitative trend is present for the other classifications.  Whereas
the overall S06 AGN fraction for galaxies in pairs (Figure
\ref{3diag}) increases by a factor of 2.5, the fraction of pairs in
which both galaxies are AGN is about 8 times higher than the control.
The fraction of double AGN expected at random can be
estimated from the square of the overall AGN fraction in a given $r_p$
bin and is shown with grey points in Figure \ref{x_jk}.  This simple
comparison shows that the observed fraction of double AGN is higher
than expected in the random case by a factor of about two and hints at
simultaneous triggering in at least some of the interacting pairs.
Liu et al. (2011b) have recently used correlated properties (such as
stellar ages, 4000 \AA\ break strength and H$\delta$ absorption
equivalent width) to infer synchronous AGN triggering.

We test for correlated AGN activity within a given pair as a
function of separation using a statistical approach.  For a given
fraction of AGN in a population, $f$, the random (i.e. uncorrelated)
fractions of pairs with zero, one or two AGN ($f_{r0}, f_{r1}, f_{r2}$
respectively) are:

\begin{equation}\label{fr0_eqn}
f_{r0} = (1-f)^2,
\end{equation}

\begin{equation}\label{fr1_eqn}
f_{r1} = 2f(1-f),
\end{equation}

\begin{equation}\label{fr2_eqn}
f_{r2} = f^2.
\end{equation}

We now assume that in a given sample with overall AGN (including QSOs)
fraction $f$, some fraction, $x$, of the pairs have correlated
emission line properties, i.e. the occurrence of an identical
classification is not random but induced by the interaction.
Therefore, the fraction of uncorrelated pairs is $(1-x)$.  In this
scheme, the correlated pairs could either be both AGN or both
star-forming.  The fraction of uncorrelated pairs with zero, one or
two AGN are then obtained by multiplying the random fractions in
equations \ref{fr0_eqn} -- \ref{fr2_eqn} by $(1-x)$.  The fraction of
correlated pairs is simply $x$ multiplied by either the AGN fraction,
$f$ (for AGN-AGN pairs) or the star-forming fraction $1-f$ (for SF-SF
pairs).  Table \ref{xeqns_tab} summarizes the equations for correlated
and uncorrelated fractions.  Note that in pairs with one AGN only the
uncorrelated case is applicable.  Finally, the actual observed
fraction of pairs with zero, one or two AGN will be the combination of
the correlated and uncorrelated fractions (i.e. the sum of the two
columns in Table \ref{xeqns_tab}):

\begin{equation}\label{f0_eqn}
f_{0} = x(1-f) + (1-x)(1-f)^2,
\end{equation}

\begin{equation}\label{f1_eqn}
f_{1} = (1-x)2f(1-f),
\end{equation}

\begin{equation}\label{f2_eqn}
f_{2} = xf + (1-x)f^2.
\end{equation}

The observed fractions of pairs with zero, one or two AGN  
($f_{0}, f_{1}, f_{2}$ respectively) and the observed total AGN fraction
($f$) in any given $r_p$ bin can therefore be combined
with any of the equations \ref{f0_eqn} -- \ref{f2_eqn} to
determine $x$.  For example, from equation \ref{f2_eqn}

\begin{equation}\label{x_eqn}
x=\frac{f_2 - f^2}{f - f^2}.
\end{equation}

\begin{center}
\begin{table}
\begin{tabular}{lcc}
\hline 
  &  Correlated fraction  & Uncorrelated fraction \\ \hline
0 AGN & $x(1-f)$ & $(1-x)(1-f)^2$ \\
1 AGN &  ... & $(1-x)2f(1-f)$ \\
2 AGN & $xf$ & $(1-x)f^2$ \\
\hline 
\end{tabular}
\caption{\label{xeqns_tab} Fraction of pairs with zero, one or two
AGN that are either correlated or uncorrelated, where $f$ is the
overall AGN fraction and $x$ is the fraction of correlated pairs. }
\end{table}
\end{center}

The fraction of pairs with correlated AGN ($xf$) is shown as a
function of $r_p$ in the lower panel of Figure \ref{x_jk}.  The errors
on $xf$ are determined with jackknife re-sampling.  In each $r_p$ bin
$xf$ is re-calculated after systematically removing the $i^{th}$ AGN
pair in turn and calculating $\delta_i = xf - x_i f_i$.  For the $N$
pairs in each $r_p$ bin the error is then $[(N-1)/N \Sigma_i
\delta_i^2]^{1/2}$.

As expected, the correlated fraction of the control sample follows the
zero line.  For the pairs, there is an increase in the fraction of
pairs with correlated double AGN as the projected separation
decreases.  Even pairs with separations of $40 < r_p < 80$ \hkpc\
exhibit a non-zero fraction of double AGN pairs that are correlated.
The raw AGN fractions in Figure \ref{frac_rp} show a strong excess of
AGN only within 40 \hkpc, with a statistically less robust (1$\sigma$)
enhancement out to 60 \hkpc.  However, the sample is likely to become
increasingly contaminated by interlopers (galaxies that are not truly
interacting) at wider separations.  An elevated (but decreasing) AGN
fraction at wide separations may therefore become increasingly
difficult to detect.  The results from the double AGN indicate that
the effect of the interaction on nuclear activity may actually be
present out to 80 \hkpc\ and beyond.  This would indicate that the AGN
phase lasts for a significant fraction of the oribital time of the
merger.  Interestingly, Patton et al. (2011) find that the central
galaxy colours of pairs remain bluer than their control sample out to similarly
wide separations.  The blue colours can be explained by elevated star
formation rates in pairs with projected separations out to 80 \hkpc\
(Scudder et al. in preparation).  It would be of great interest to
extend star formation and AGN analyses to even wider separations.

\section{Discussion}\label{comp_sec}

\subsection{Comparison to previous work}

As reviewed in the Introduction, there has been considerable
disagreement in the literature regarding the connection between galaxy
mergers and AGN activity.  The two main techniques that have
previously been used to tackle this connection are 1) through the
imaging of AGN host galaxies (to search for morphological
disturbances) and 2) the presence of AGN in merger candidates.  The
former technique has the obvious difficulty of imaging depth and the
ability to detect morphological disturbances.  These issues have been
discussed in the literature and we refer the reader to the papers
focused explicitly on this technique for more details (e.g. Ramos
Almeida et al. 2011a,b).  In this paper, we have taken the second
approach: to quantify the AGN fraction in a sample of likely merger
candidates.  This technique also has its controversies.  Even with
large samples and carefully matched controls, a disagreement remains
as to whether close pairs have a higher AGN fraction (Li et al. 2006,
2008; Alonso et al. 2007; Woods \& Geller 2007; Ellison et al. 2008.
Darg et al. 2009; Rogers et al. 2009).

Our results can be used to shed light on some of
the reasons behind the discrepant results in the literature, as well
as some of the selection effects which do \textit{not} seem to play a
major role.  For example, the choice of mass ratio has a relatively
minor effect on the detection of an enhanced AGN fraction.  We find
that restricting our analysis to approximately equal mass mergers
produces quite similar results to the full 1:10 mass range (Figure
\ref{3mratio}).  The result is also robust to the choice of
relative velocity: strong enhancements in the AGN fraction are still
seen at small separations even with a relatively relaxed threshold of
$\Delta V < 500$ \kms.  The choice of AGN diagnostic is also not a
major factor; we see increasing AGN fractions at smaller projected
separations for all three of the standard BPT diagnostics (Figure
\ref{3diag}).  

One selection criterion that does play a significant role is the S/N
imposed on the detection of emission lines. The S/N$>$5 cut used in
our work is fairly typical of emission line studies wishing to
accurately determine gas-phase metallicities (as reviewed by Kewley \&
Ellison 2008).  However, the location of a galaxy on the BPT diagram
may be less sensitive to S/N ratio.  Moreover, some authors have
argued that imposing S/N cuts biases the sample against low luminosity
AGN which preferentially have LINER-like spectra (e.g. Woods \& Geller
2007; Cid-Ferandes et al. 2010).  We confirm this general trend by
calculating the fraction of LINERs in the full SDSS sample as a function
of S/N.  Although LINERs lie in a slightly different part of the BPT
diagram (e.g. Figure 3 of Cid-Fernandes et al. 2010) we classify
them using the [OI] diagnostic of Kewley et al. (2006b).  The LINER
fraction rises from 4 per cent for a S/N$>$ 5 to 17 per cent for a
S/N$>$1.  However, the fraction of AGN in the pairs sample is actually
lower for lower S/N cuts.  For example, at the closest separations
($r_p < 10$ \hkpc), the AGN excess is only 20 per cent for a cut
of S/N$>2$.  When the criterion is increased to S/N$>$ 8, the AGN
excess is a factor of 3.5 (recall that for our fiducial S/N$>$ 5
cut the enhancement is a factor of 2.5).  This trend of higher AGN
fraction for more stringent S/N criteria indicates that AGN in
the pairs sample are \textit{not} dominated by objects with LINER-like
spectra.

One important component of our analysis has been a careful treatment
of selection effects.  As discussed in Section \ref{sample_sec},
incompleteness at small separations in the SDSS leads to a biased
distribution of masses as a function of projected separation.  The AGN
fraction is extremely sensitive to galactic stellar mass: for
M$_{\star} = 10^{10}$ M$_{\odot}$ the fraction of emission line
galaxies classified as AGN by S06 is $\sim$ 20\%, rising to about 75\%
by M$_{\star} = 3 \times 10^{10}$ M$_{\odot}$.  There are similarly
steep dependences of the AGN fraction when using the K01 and K03
diagnostics, although the overall fraction is lower than with
S06. Therefore, even a 0.1 dex bias in mass can change the fraction of
emission line galaxies classified as AGN by 10\%.  We have mitigated
this effect through a cull of our wide pairs (to remove mass dependences on
projected separation) and a careful matching of a control sample (to
facilitate a relative comparison).

The enhanced AGN fraction at close separations is striking when
displayed as a function of projected separation (Figure
\ref{frac_rp}).  The significance of our result would be diminished if
we considered the AGN fraction of the pairs sample as a whole (i.e.
not as a function of $r_p$) and
compared it to the control.  Our work has therefore benefitted from a
large sample where previous works have calculated AGN fractions for
pairs and controls in a wholesale way (e.g. Alonso et al. 2007; Woods
\& Geller 2007; Ellison et al. 2008).  It is also worth noting that
our pairs sample differs from these previous works, in that it
includes galaxies without strong emission lines.  The AGN fractions
quoted here are fractions of all galaxies, not fractions of emission
line galaxies that are classified as AGN.  Repeating our analysis for
an emission line only sample shows a much more modest increase in the
AGN fraction at small separations, by only about 20\% (using the S06
diagnostic).  As demonstrated by Figure \ref{tran_frac}, the overall
fraction of galaxies with emission lines (star-forming, composite and AGN)
increases in pairs with smaller projected separations.  Calculating
AGN fractions for a complete galaxy sample therefore accounts for the
migration of quiescent galaxies into the emission line regime.  Our
technique therefore tracks a single sample of galaxies and their
changing emission line properties.  For an emission line only sample,
a changing AGN fraction as a function of $r_p$ is then a combination
of the changing size emission line sample, plus the changing fraction
of AGN.

\subsection{Multiple AGN channels?}

The arguments above help to reconcile the disparate results in the AGN
fraction of close pairs by demonstrating some of the selection effects
that can lead to different AGN fractions.  However, our conclusion
that mergers can lead to AGN should be distinguished from whether
\textit{all} AGN are the result of mergers.  In theory, any process
that can funnel gas to the centre of a galaxy may be expected to
increase the accretion rate of the black hole and lead to an AGN.
Indeed, there is an observational link between enhanced central star
formation rates and black hole activity that extends beyond mergers to
the general galaxy population (Cid-Fernandes et al. 2001; Kauffmann et
al. 2003a; Silverman et al. 2009; Reichard et al. 2009).  Other
evidence has led to claims that mergers are indeed unlikely to be the
sole AGN trigger (e.g. Li et al. 2006; Lutz et al. 2010; Mullaney et
al.  2011), which may play a role in explaining why many imaging
surveys of AGN hosts detect no signs of (an excess of) recent merger
activity (e.g. Grogin et al. 2005; Gabor et al. 2009; Cisternas et al
2011; Kocevski et al. 2011).  Two channels for AGN activity are often
recognised in simulations, the violent, merger-induced `quasar-mode'
and the more gentle `maintenance-mode' (e.g. Di Matteo et al. 2005;
Hopkins \& Hernquist 2006; Hopkins et al. 2008).  There is general
observational support for the association of quasars and radio
galaxies with mergers (e.g. Canalizo \& Stockton 2001; Ramos Almeida
et al. 2011a,b and references therein) when suitable control samples
are used.  However, the work presented here is perhaps more surprising
(and challenges the simple two channel model described above) in that
we find an excess of relatively low luminosity (mostly Seyfert-like)
AGN in interacting pairs.

If there are multiple pathways to creating an AGN, we should
distinguish between asking `what fraction of close pairs have AGN?'
(the question addressed in the main body of this paper) and `what
fraction of AGN are associated with mergers?'.  The latter question
epitomizes the approach of most imaging surveys that have not found a
significant difference between the disturbed, or close pair, fractions
in AGN and non-AGN samples.  We can also ask this second question of
our SDSS sample.  The same AGN classification criteria described in
Section \ref{diag_sec} are applied to the complete SDSS main galaxy
catalogue.  Of the 43,436 galaxies that are classified as AGN,
0.030$\pm$0.001 have a close companion within 200 \kms\ and 30 \hkpc.
We also define a sample of 43,436 non-AGN galaxies; the best
simultaneous match in stellar mass and redshift to each of the AGN
galaxies.  The fraction of non-AGN with a close companion within 200
\kms\ and 30 \hkpc\ is 0.020$\pm$0.001 (the error is based on simple
poisson statistics)\footnote{The fraction of galaxies with a close
companion in both the AGN and non-AGN samples is not accurate in the
absolute sense, due to spectroscopic incompleteness and other selection
biases.  For this simple comparison, we have not attempted an accurate
correction, but assume that both the AGN and non-AGN are affected
identically and hence the fraction of AGN and non-AGN with a close
companion can be compared in a relative sense.}.  Therefore, from our
sample of pairs we can say that close interactions result in a higher
AGN fraction \textit{and} AGN in the full sample show an excess of
close companions.  This contrasts with the imaging studies (mostly at
higher redshift) listed above which do not find an excess of mergers
associated with X-ray selected AGN.  However, even accounting for the
spectroscopic incompleteness (Patton \& Atfield 2008), it is clear
that not all AGN in the SDSS sample have a close companion.

If mergers are not the dominant cause of AGN activity, then the next
goal is to identify the main mechanism that leads to black hole
accretion.  Simulations provide some insight here, by investigating
the relative efficiencies of direct `cold flow' gas accretion and
mergers (e.g. Hopkins \& Hernquist 2006; Di Matteo et al. 2011).
Indeed, Bournaud et al. (2011) argue that, at least at high redshift
where galactic gas fractions are high due to inflows from cold
streams, disk instabilities could lead to significant AGN fuelling.  This
argument is supported by the prevalence of clumpy disks in high
redshift galaxies (Elmegreen et al. 2007).  However, Wisnioski et
al. (2011) do not find any AGN dominated disks in their small sample
of clumpy $z \sim 1$ disks.  Alternatively, Ellison et al. (2011) have
recently demonstrated that barred galaxies with log M$_{\star} >
10^{10}$ M$_{\odot}$ have SFRs that are enhanced at a similar level as
close pairs.  However, due to their relative ubiquity, the enhanced
SFRs in barred galaxies contribute at least three times more to the
central stellar mass build-up than mergers.  We will investigate the
AGN fractions in barred galaxies in a forthcoming paper (Nair et al.
in preparation).

\section{Conclusions}

We have compiled a large (11,060) sample of close galaxy pairs,
selected with careful attention to the photometry at small
separations, emission line quality and biases due to small separation
incompleteness. The AGN fraction of the pairs sample is compared
to a mass- and redshift-matched control sample with 10 controls per
pair galaxy.

The main conclusions of this paper are:

\begin{enumerate}

\item   The AGN fraction in close pairs of galaxies increases by a factor
of up to $\sim$ 2.5 at projected separations $r_p < 10$ \hkpc. Although
the increase depends slightly on the choice of AGN diagnostic,
elevated AGN fractions are found for closely separated pairs 
independent of the diagnostic choice.  

\item The excess of AGN is larger for more stringent S/N cuts,
indicating that the AGN in pairs are not dominated by objects with
LINER-like spectra.

\item We discuss how AGN are likely to be triggered by multiple
processes, of which mergers are just one channel.  However,
in contrast to models, our results show that mergers can result in
an increase in relatively low luminosity AGN, as well as the powerful
radio galaxies and quasars that have been investigated by other studies.

\item  The fraction of AGN in close pairs depends mildly on the ratio
of stellar masses.  The highest enhancements in AGN fraction are
seen for the equal mass pairs.  The lower mass galaxies in unequal
mass pairs show relatively little enhancement AGN fraction.

\item In addition to an increase in the `pure' AGN fraction at decreasing
projected separations, both the star-forming and composite fractions
also increase.  Increases in all three emission line classes indicates
that some quiescent galaxies have been transformed during the interaction
process.

\item The fraction of pairs in which \textit{both} galaxies are
 classified as AGN by the Stasinska et al. (2006) AGN diagnostic is a
 factor of up to 8 higher than a sample of `control pairs'.  This
 exceeds the expected random fraction of double AGN in pairs by around
 a factor of two, indicative of correlated AGN triggering between
 the companions.

\item The correlated double AGN fraction is investigated using a statistical
approach and found to increase at small separations, but remains 
elevated out to 80 \hkpc.  This is evidence that the AGN phenomenon
may persist in relatively wide separation pairs.

\end{enumerate}

The results of this paper are in clear support of interaction-driven
AGN activity that occurs well before final galaxy-galaxy coalescence.
Since the merger timescale may be as long as a gigayear (Kitzbichler
\& White 2008), this does not provide a strong constraint on a delay
between triggered star formation and AGN activity.  However, the shape
of the double AGN fraction as a function of projected separation (Figure
\ref{x_jk}) is very similar to the colour offsets (Patton et al. 2011) and
star formation rate enhancements (Scudder et al.  in preparation)
in our sample.  Scudder et
al. compare the observed star formation rates with simulations and
demonstrate that they can be reproduced with a burst of star formation
occurring after the first pericentric passage, followed by a more
significant burst at final coalescence.  The similarity between star
formation and AGN enhancements as a function of projected separation
may be indicative of a similar process for accretion onto the nucleus
with a relatively small time lag.  Delays of a few hundred million
years (Wild et al.  2010; Hopkins 2011) between peaks in star
formation and accretion rate correspond to changes in projected
separation of order the bin width in our projected separation figures
(depending on relative velocity and the orientation).  Megayear delays
are therefore relatively prompt compared with the resolution provided
by the observations presented here.  The high fraction of composite
galaxies also demonstrates that the AGN and star formation processes
are on-going concurrently, with no need to shut down the latter before
the former begins (c.f. Li et al. 2008; Schawinski et al. 2009).

\section*{Acknowledgments} 

We are grateful to the MPA/JHU group for
access to their data products and catalogues (maintained by Jarle
Brinchmann at http://www.mpa-garching.mpg.de/SDSS/). 
SLE and DRP acknowledge the receipt of an NSERC Discovery grant which
funded this research.  Frederic Bournaud and Grazyna Stasinska
provided insightful comments the manuscript.

    Funding for the SDSS and SDSS-II has been provided by the Alfred
    P. Sloan Foundation, the Participating Institutions, the National
    Science Foundation, the U.S. Department of Energy, the National
    Aeronautics and Space Administration, the Japanese Monbukagakusho,
    the Max Planck Society, and the Higher Education Funding Council
    for England. The SDSS Web Site is http://www.sdss.org/.

    The SDSS is managed by the Astrophysical Research Consortium for
    the Participating Institutions. The Participating Institutions are
    the American Museum of Natural History, Astrophysical Institute
    Potsdam, University of Basel, University of Cambridge, Case
    Western Reserve University, University of Chicago, Drexel
    University, Fermilab, the Institute for Advanced Study, the Japan
    Participation Group, Johns Hopkins University, the Joint Institute
    for Nuclear Astrophysics, the Kavli Institute for Particle
    Astrophysics and Cosmology, the Korean Scientist Group, the
    Chinese Academy of Sciences (LAMOST), Los Alamos National
    Laboratory, the Max-Planck-Institute for Astronomy (MPIA), the
    Max-Planck-Institute for Astrophysics (MPA), New Mexico State
    University, Ohio State University, University of Pittsburgh,
    University of Portsmouth, Princeton University, the United States
    Naval Observatory, and the University of Washington.

\end{document}